\documentclass[twocolumn]{aastex631}

\shorttitle{Dynamical Interactions and Mass Loss Within the Uranian
  System}
\shortauthors{Stephen R. Kane \& Zhexing Li}

\begin{document}

\title{Dynamical Interactions and Mass Loss Within the Uranian System}

\author[0000-0002-7084-0529]{Stephen R. Kane}
\affiliation{Department of Earth and Planetary Sciences, University of
  California, Riverside, CA 92521, USA}
\email{skane@ucr.edu}

\author[0000-0002-4860-7667]{Zhexing Li}
\affiliation{Department of Earth and Planetary Sciences, University of
  California, Riverside, CA 92521, USA}


\begin{abstract}

The origin and evolution of planetary rings and moons remains an
active area of study, particularly as they relate to the impact
history and volatile inventory of the outer solar system. The Uranian
system contains a complex system of rings that are coplanar with the
highly inclined planetary equator relative to the orbital
plane. Uranus also harbors five primary regular moons that play an
important role in the distribution of material that surrounds the
planet. Here we present the results of a dynamical simulation suite
for the Uranian system, intended to explore the interaction between
the five primary regular moons and particles within the system. We
identify regions of extreme mass loss within 40 planetary radii of
Uranus, including eccentricity excitation of particle orbits at
resonance locations that can promote moonlet formation within the
rings. We calculate a total dynamical particle mass loss rate of 35\%
within $0.5 \times 10^6$~years, and 40\% mass loss within
$10^7$~years. We discuss the implications for post-impact material,
including dynamical truncation of stable ring locations, and/or
locations of moon formation promoted by dynamical excitation of ring
material.

\end{abstract}



\section{Introduction}
\label{intro}

The giant planets of the solar system each present opportunities to
study the complex interaction between their gravitational influence
and the numerous bodies that lie within their Hill spheres. Notable
features within each giant planet system, such as rings and moons,
provide traceable elements of the dynamical history and angular
momentum transfer within those systems, including evidence of past
impacts, collisions, and ejections. The numerous particles that exist
within the present ring systems of Jupiter
\citep{showalter1987,porco2003}, Saturn
\citep{pollack1975d,porco2005a}, Uranus \citep{elliot1977c,tyler1986},
and Neptune \citep{lane1989,showalter2020} are particularly rich
sources of dynamical histories within those planetary
environments. For example, Saturn's rings have been studied
extensively with respect to their dynamics
\citep{goldreich1978b,bridges1984}, velocity dispersion
\citep{salo1995}, and the use of density waves to infer fundamental
properties of the planetary interior
\citep{hedman2013c,mankovich2019,mankovich2021}. Planetary ring
material have a variety of sources, including collision events or the
desiccation of moons and Kuiper belt objects by tidal forces
\citep{canup2010,hyodo2017a,hyodo2017b}. The presence and
sustainability of ring material is also an intricate function of the
architecture of planetary moons, as well as the intrinsic properties
of the planet itself
\citep{petit1988c,rubincam2006,nakajima2020,kane2022c}. Furthermore,
the vast number of bodies orbiting the solar system outer planets has
motivated numerous searches for exomoons
\citep[e.g.,][]{hinkel2013b,kipping2013d,heller2014c,hill2018} and
rings
\citep[e.g.,][]{arnold2004c,kenworthy2015b,zuluaga2015a,sucerquia2020b}.
The solar system giant planets and their associated companions can
serve as important analogs for compact exoplanetary systems, revealing
insight into their formation and architectures
\citep{kane2013e,makarov2018,dobos2019,batygin2020b}.

Ice giant planets are of particular interest with respect to their
dynamical environment since they have played a significant role in
planet formation and evolution at the outer edges of the
protoplanetary disk \citep{ford2007a,dawson2012a}. The Neptunian
system may be atypical of ice giant moon systems given the relatively
rare capture scenario for its major moon, Triton
\citep{agnor2006,li2020d,li2020f}. The Uranian system is especially
diverse with respect to its dynamical history and system of moons and
rings, providing a valuable template from which to explore the
development of ice giant architectures
\citep{peale1999a,esposito2002,jacobson2014f}. The Uranian moons
continue to be the target of observations that aim to update their
orbital ephemerides and study their compositions
\citep{brozovic2022,paradis2023}. Uranus has five primary regular
moons, all of which lie within 25 planetary radii of the planet and
exhibit a myriad of geological features
\citep{camargo2022,kirchoff2022,castillorogez2023}. The dynamical
state of the moon system has also been studied in detail
\citep{dermott1988c,cuk2020b}, including long-term stability issues
and collision scenarios \citep{cuk2022b}, tidal evolution
\citep{tittemore1988,tittemore1989b,tittemore1990a}, and resonances
\citep{lazzaro1984,quillen2014c,french2015b,charalambous2022a}. The
Uranian system contains substantial evidence of a rich impact
history. The most compelling evidence is contained within Uranus
itself, whose large axial tilt origin has been explained via various
mechanisms \citep{boue2010,lu2022,saillenfest2022}, and frequently
attributed to a giant impactor early in its history
\citep{korycansky1990,slattery1992,kegerreis2018}. Such evidence is
also contained upon the moon surfaces, exhibiting a vast range of
geological and topographical features
\citep{johnson1987c,johnson1987g}. Indeed, the formation and
subsequent evolution of the moons may be attributed to a giant impact
event \citep{izidoro2015b,ida2020a,chau2021,salmon2022a,woo2022b}, and
is consistent with the moon's prograde motion and relative orbital
coplanarity with the planet's equatorial plane
\citep{morbidelli2012b}. Uranus also has a significant ring system
which, when including the relatively tenuous outer ring system
\citep{showalter2006a}, extends to four planetary radii and well
beyond the fluid body Roche limit. The Uranian rings have long been a
discussion topic regarding their formation and sustainability
\citep{goldreich1979b,esposito1989}, and suggested as a probe of the
planetary interior via ring seismology \citep{ahearn2022}. The rings
also possess a complex relationship with the moons, notably with
accretion processes and moonlets within the Roche limit
\citep{canup1995}, and the kinematics of shepherding interactions
between the moons and rings \citep{goldreich1987,porco1987a}. An
exploration of the concise dynamical influence of the moons over the
distribution of past and present material within the Uranian system
will provide further insight into the evolution of this fascinating
system of moons and rings.

Here, we present the results of a detailed dynamical analysis for the
five primary regular moons in the Uranian system, and their influence
on injected particles into the system. The results show the dynamical
exclusion regions and ring material mass loss rates around Uranus for
timescales of up to $10^7$~years. Section~\ref{arch} describes the
architecture of the Uranian ring and moon system, with comparison to
those of Jupiter and Saturn. Section~\ref{dynamics} provides a
description of the dynamical simulation methodology and the results
for the moons and injected particles. The simulation results and their
implication for moon and ring evolution are discussed in
Section~\ref{discussion}. Concluding remarks and suggestions for
additional work are provided in Section~\ref{conclusions}.


\section{Architecture of the Uranian System}
\label{arch}

Amongst the known population of moons, regular moons of giant planets
are particularly notable in that they likely formed either with the
planet or via subsequent collision events
\citep{lunine1982a,canup2002,canup2006,ronnet2020}, as evidenced by
their typically equatorial prograde orbits, and are often large enough
to exhibit hydrostatic equilibrium, resulting in a near-round
morphology. At the time of writing, the Uranus system is known to
harbor a total of 27 moons, including the five major regular moons of
(in order of increasing semi-major axis) Miranda, Ariel, Umbriel,
Titania, and Oberon\footnote{\tt
  https://solarsystem.nasa.gov/moons/overview/}. The smallest of these
moons, Miranda, has a mean radius of $\sim$235~kms, placing it near
the limit of hydrostatic equilibrium
\citep{thomas1988a,beddingfield2015a}. Even so, Miranda is almost
seven times more massive than the combined mass of the smaller moons
within the system. Miranda, in turn, is $\sim$20 times less massive
than the next largest moon, Ariel. This demonstrates the clear
gravitational dominance of the five primary moons, and the relatively
small influence of Miranda compared with the other four.

\begin{figure*}
  \begin{center}
    \includegraphics[angle=270,width=16.0cm]{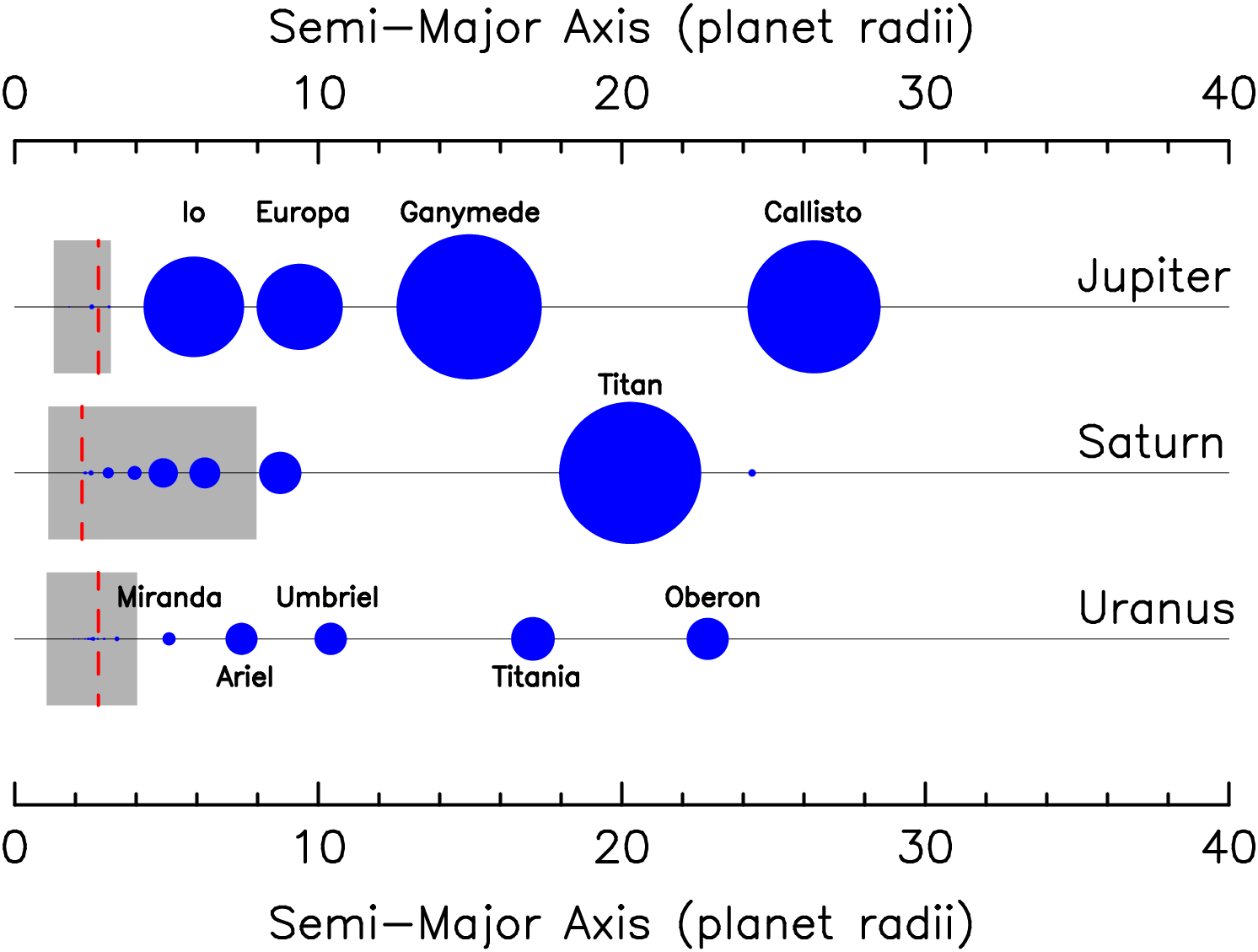}
  \end{center}
  \caption{The regular moons and rings of the Jupiter (top), Saturn
    (middle), and Uranus (bottom) systems. The relative sizes of the
    moons are shown, and their semi-major axes are provided in units
    of the host planet radii. The extent of the current ring systems,
    including dense and tenuous rings, are shown as gray regions, and
    the vertical red dashed lines indicate the location of the fluid
    satellite Roche limit for each planet.}
  \label{fig:radii}
\end{figure*}

Shown in Figure~\ref{fig:radii} is a scaled view of the moon systems
for Jupiter (top), Saturn (middle), and Uranus (bottom), where the
separations from the planetary centers are in units of the host planet
radius. The regular moons (larger moons that likely formed in orbit of
the planet) within each system are shown in blue and their sizes are
scaled relative to each other. The location of the fluid satellite
Roche limits (red dashed lines) are 2.76, 2.22, and 2.75 planetary
radii for Jupiter, Saturn, and Uranus, respectively. Also shown are
the extent of the ring systems for each planet (gray regions). Note
that the shaded ring regions incorporate both dense and tenuous rings,
including faint outer ring systems. For example, the indicated ring
region for Jupiter extends to the Thebe gossamer ring (3.16 Jupiter
radii), and the D--E rings are included for Saturn, extending to 7.96
Saturn radii. In the case of Uranus, the faint $\mu$ and $\nu$ rings,
discovered using observations by the Hubble Space Telescope (HST)
\citep{showalter2006a} extend out to 4.03 Uranus radii.

Although there are significant differences in the size distribution
and architectures of the three planet/moon systems represented in
Figure~\ref{fig:radii}, there are also numerous similarities. All
three planets harbor their regular moons within 30 planetary radii of
their host planet, and within $1/25$ of their respective Hill radii.
The sizes and masses of the largest Uranian moons with respect to
Uranus are roughly proportional to the relative sizes and masses of
the Galilean moons compared with Jupiter, consistent with a scaling of
moon formation with the mass of the primary planet
\citep{canup2006}. In addition to the gravitational effect of the
orbiting moons on material surrounding the planet
\citep{petit1988c,nakajima2020}, such as the Laplace resonance effects
of the Galilean moons \citep{malhotra1991,peale2002b,kane2022c}, there
are various other factors at work that both contribute to and remove
such material \citep{daisaka2001}. Additions to material may originate
from impacts and degassing events on moons
\citep{esposito2002}. Beyond the Roche limit, material will often
accrete to form new moons, such as some of those moons that are
currently present within Saturn's ring structure
\citep{charnoz2010,crida2012,salmon2017}. Removal of ring material may
involve several non-gravitational effects, particularly for tenuous
rings that are more susceptible to such processes as
Poynting-Robertson drag, the Yarkovsky effect, and those related to
planetary magnetospheres
\citep{burns1999,rubincam2006,kobayashi2009a}. The dynamical
simulations described in this work focus on dense material that is
more likely to be influenced by the gravitational and accretion
effects imposed by the planet and moons in the system.


\section{Dynamical Simulations}
\label{dynamics}

This section describes the dynamical simulations, their configuration,
results for the five major moons, and particle injections.


\subsection{Dynamics of the Uranian Moons}
\label{moons}

The dynamical evolution of the Uranus moons have previously been
studied in detail \citep{greenberg1975c,greenberg1975f,lainey2008},
including their resonances and the consequences for their tidal
development \citep{cuk2020b,charalambous2022a}. Our simulations
explore the dynamical influence of the five main regular moons of
Uranus since, as described in Section~\ref{arch}, even the smallest of
these moons (Miranda) is substantially more massive than the combined
mass of the smaller moons, and so these five major moons dominate the
gravitational perturbations within the Uranian Hill sphere. Therefore,
we commenced our dynamical analysis by examining the orbital stability
of the major moons over 1--10 million year timescales. The purpose of
these initial simulations were to establish a baseline performance of
our methodology and examine the eccenetricity variations of the major
moons.

\begin{figure*}
  \begin{center}
    \includegraphics[angle=270,width=16.0cm]{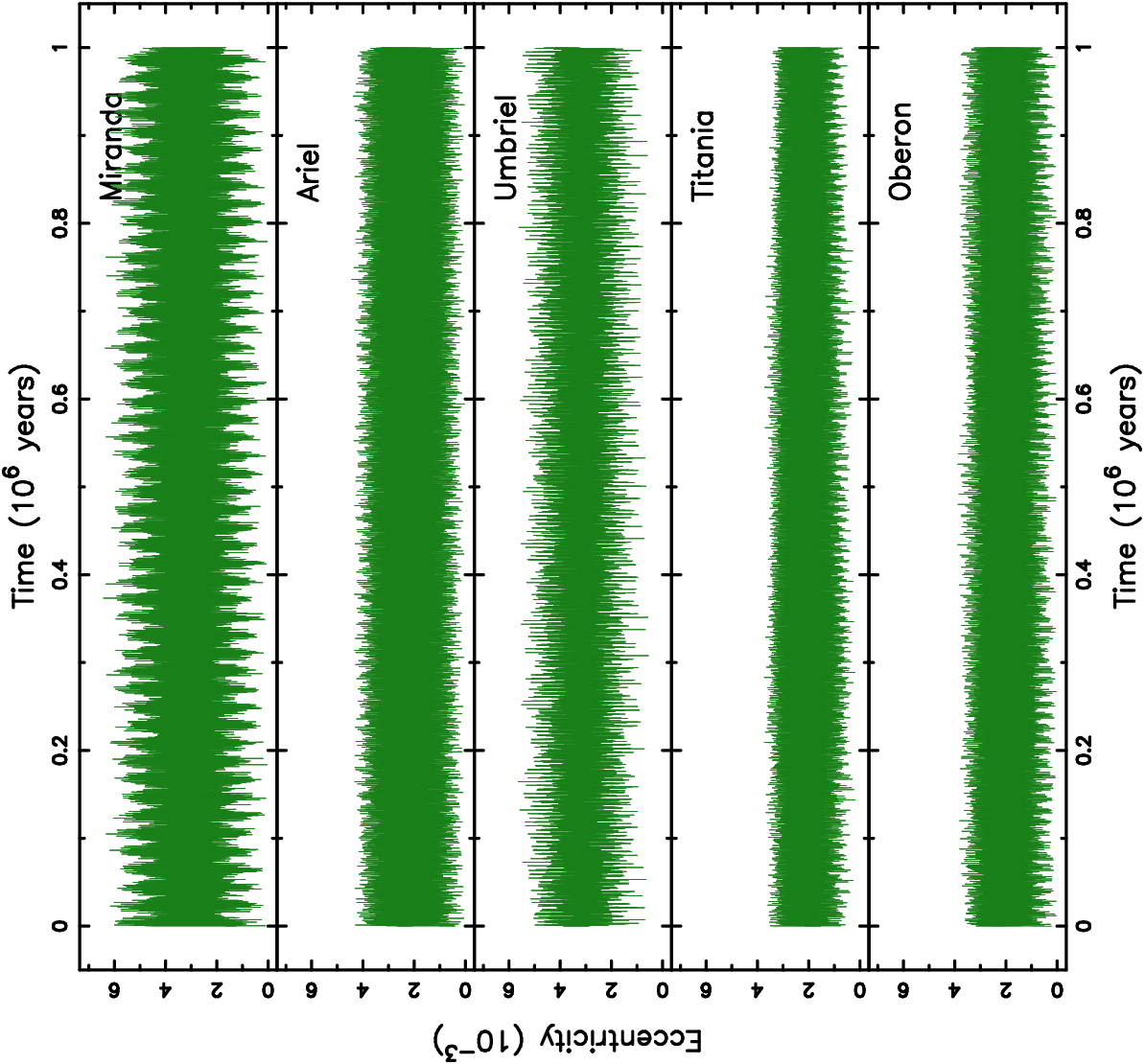}
  \end{center}
  \caption{Eccentricity as a function of time for the Uranus major
    moons of Miranda, Ariel, Umbriel, Titania, and Oberon. The
    eccentricity values were recorded every 100 years during a
    $10^6$~year simulation, as described in Section~\ref{dynamics}.}
  \label{fig:ecc}
\end{figure*}

Our simulations were conducted using the REBOUND N-body integrator
package \citep{rein2012a} that applies the symplectic integrator
WHFast \citep{rein2015c}. The initial conditions and orbital elements
of the Uranian moon system were extracted from the Jet Propulsion
Laboratory (JPL) Planetary and Lunar Ephemerides DE440 and DE441
\citep{park2021}. These data were cross-referenced with those provided
by Scott Sheppard \footnote{\tt
  https://sites.google.com/carnegiescience.edu/sheppard/}.  The
simulations do not incorporate the 2.3\% oblateness of Uranus
\citep{helled2010b}, compared with the 6.5\% oblateness of Jupiter
\citep{buccino2020a}, nor do they include the effects of tidal
dissipation. Note that the moon system stability has previously been
verified for non-MMR-crossing conditions
\citep[e.g.,][]{dermott1986,laskar1986b,malhotra1989,lazzaro1991}.
The resulting predicted eccentricity evolution of the moons over a
period of $10^6$~years are shown in Figure~\ref{fig:ecc}. Over the
timescale shown, the orbits of the moons are remarkably stable, with
eccentricity variations generally remaining below 0.005. The exception
to this is the eccentricity of Miranda, which regularly rises above
0.005 and exhibits several modes of variability frequency. A fourier
analysis of the eccentricity data for Miranda reveals a high frequency
variation with a period of $\sim$475~years, similar to the other major
moons, in addition to a low frequency eccentricity variation with a
period of $\sim$20,470~years. These high frequency results are
consistent with those of \citet{malhotra1989}, whereas the duration of
their integration was not sensitive to the low frequency variations
found in our data. However, there are some important caveats to note
regarding the orbital evolution of the system. Mean motion resonances
(MMR) play a critical role in the evolution of the orbits, as they are
relatively efficient in transferring angular momentum between bodies
that can lead to eccentricity increases and tidal effects. Although
none of the major moons are currently in MMR with each other, recent
simulations have demonstrated a past Ariel--Umbriel 5:3 MMR and
Miranda--Umbriel 5:3 MMR \citep{tittemore1990a,cuk2020b}. Such
resonances may have had a particularly profound effect on both the
orbit and geology of Miranda
\citep{hammond2014b,beddingfield2015a,beddingfield2022b}, whereby
inclination and eccentricity excitation of Miranda's orbit resulted in
significant tidal heating of the moon
\citep{dermott1988c,cuk2020b}. Furthermore, the dynamical simulations
performed by \citet{cuk2020b} reveal the potential for an
Ariel--Umbriel 5:3 MMR crossing event beyond $10^7$~years that excites
the eccentricity of Miranda to $\sim$0.03. Thus, MMR events will
continue to play a role in the evolution of the Uranian moon
system. Indeed, locations of MMR will be a crucial factor in
understanding the effect of the major moons on other material that
lies within the Uranian system, as described in
Section~\ref{influence}.


\subsection{Gravitational Influence of the Major Moons}
\label{influence}

To investigate the gravitational influence of the major moons on
particles within the system, most particularly material that may
potentially participate in the formation of rings or moonlets, we
conducted an extended series of dynamical simulations. These
simulations utilized the same REBOUND framework described in
Section~\ref{moons} and introduced particles into the system. We
adopted particle densities equivalent to water ice (0.917~g/cm$^3$),
and spherical radii of $\sim$1~meter, yielding a particle mass of
k$\sim$3841~kg. The particles were injected into circular orbits that
are coplanar with the Uranus equator. The particles were injected into
1000 evenly spaced semi-major axis locations between 1 and 40
planetary radii, resulting in a distance resolution of
$\sim$1000~kms. The inner boundary for the semi-axis range was chosen
to ensure that our simulations explore the region interior to the
fluid satellite Roche limit of Uranus, which is $\sim$2.75 planetary
radii (see Section~\ref{arch}). The outer boundary of 40 planetary
radii was chosen in the context of the planet--moon separations shown
in Figure~\ref{fig:radii}. The Hill radius of Uranus is immense (more
than 2600 planetary radii) due to its relatively large semi-major
axis, but 40 planetary radii sufficiently captures the extent of the
major moons. The duration of each simulation was $10^7$ years, which
translates to $1.2\times10^8$ orbits at the outer boundary of our
chosen semi-major axis range. In order to adequately sample the
dynamical interactions within the system, the time step for the
simulations was 0.05 of the orbital period of Miranda, except for
locations interior to Miranda's orbit where the time step was 0.05 of
the particle orbital period. The evolution of the orbital parameters
for all objects within the system were output every 100 years to allow
analysis or individual orbital elements, such as the eccentricity
evolution of the major moons described in Section~\ref{moons}. The
survival of the injected particles at each location were evaluated
based on the fraction of the simulation time for which they remained
in orbit within the system. Removal from the system could result from
either ejection from the Uranus gravity well or collision with one of
the other bodies included in the simulation.

\begin{figure*}
  \begin{center}
    \includegraphics[angle=270,width=16.0cm]{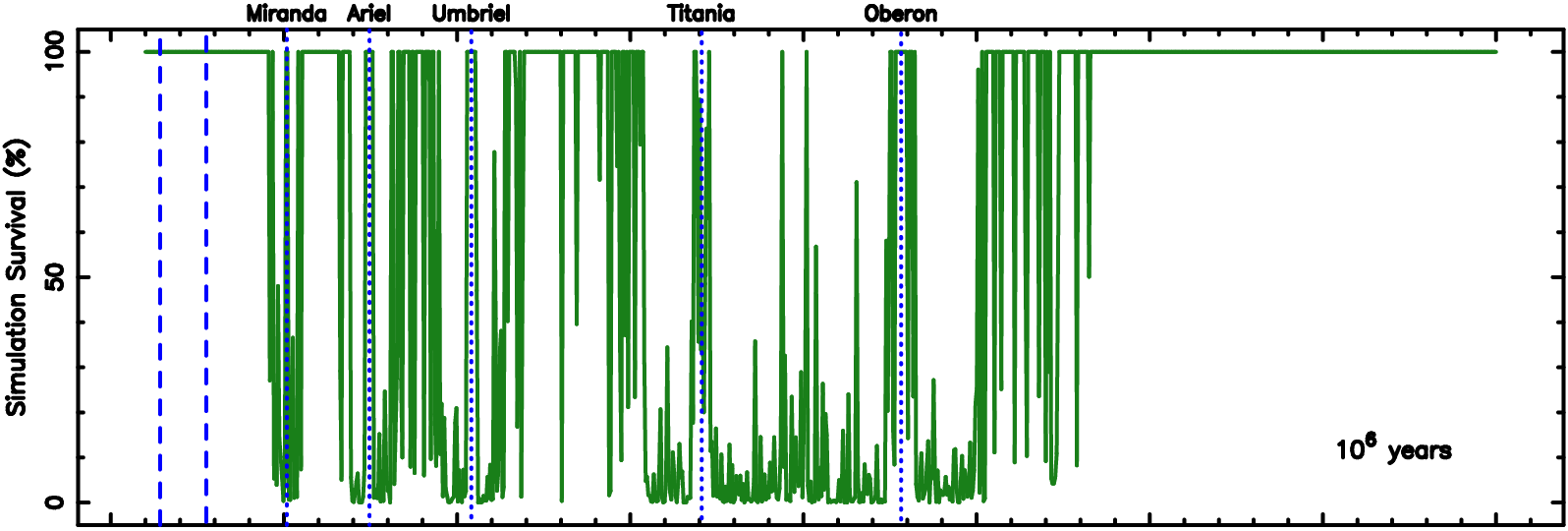} \\
    \includegraphics[angle=270,width=16.0cm]{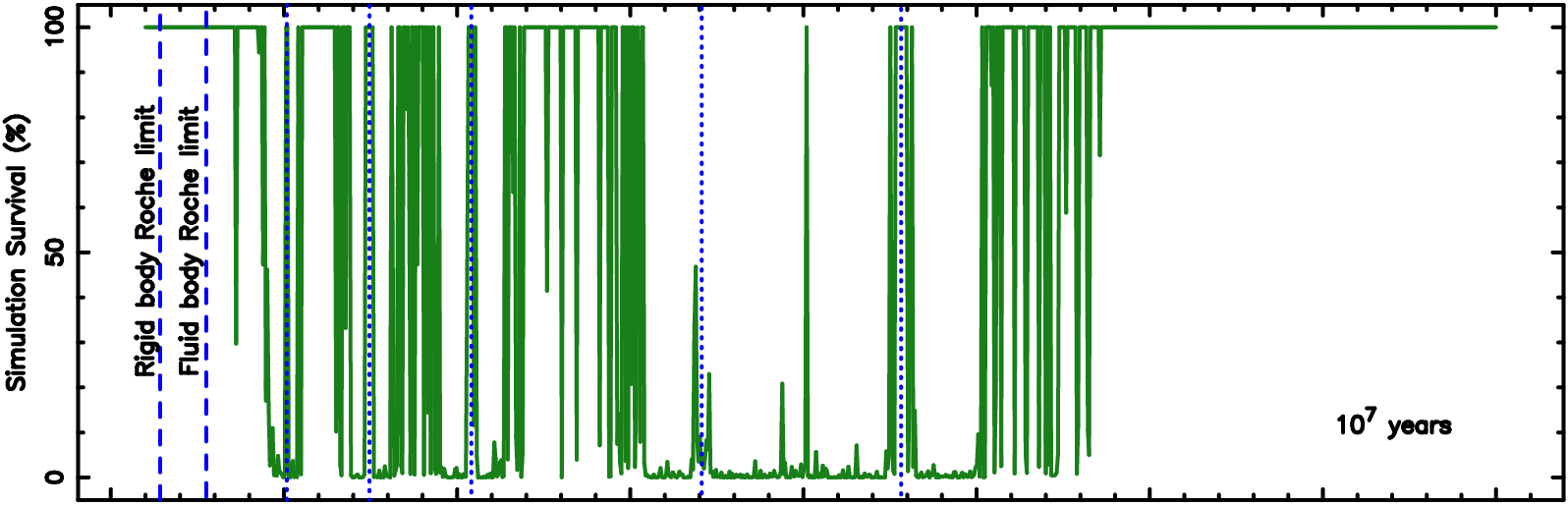} \\
    \includegraphics[angle=270,width=16.0cm]{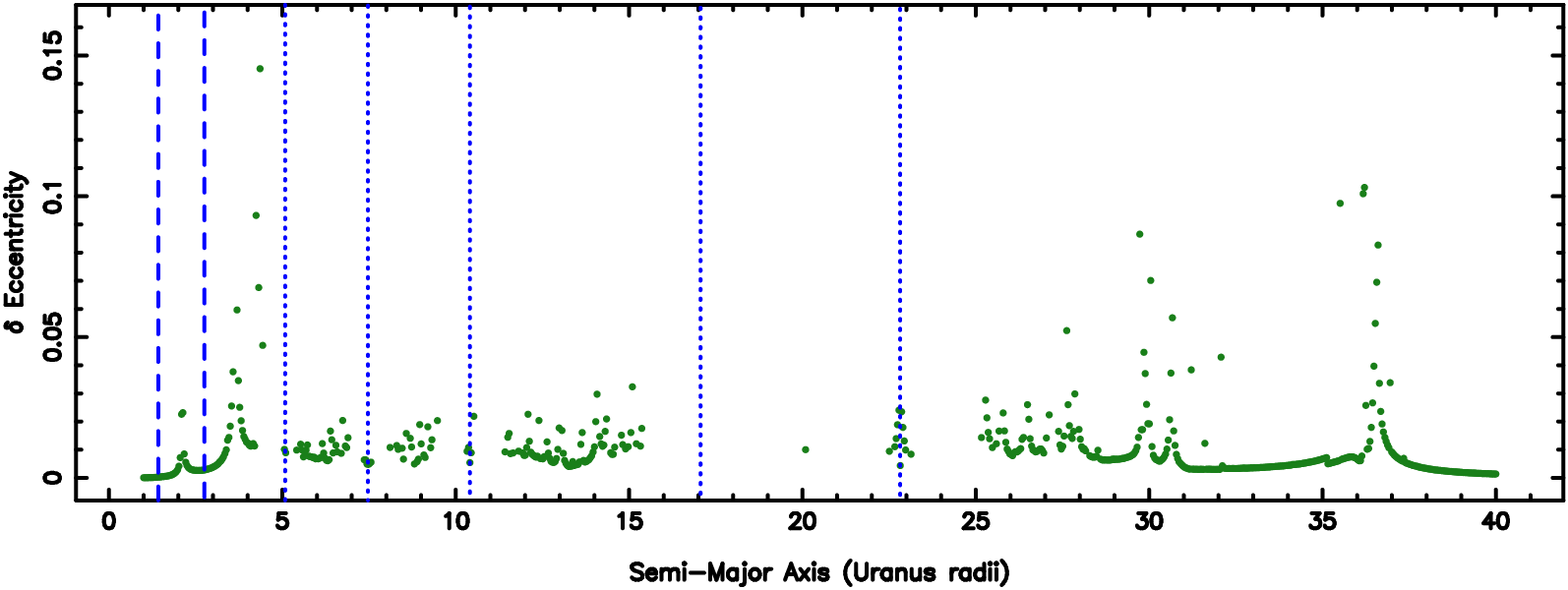}
    \end{center}
  \caption{Results of the particle injection and survival for the
    dynamical simulations, showing the outcome after $10^6$ years (top
    panel) and $10^7$ years (middle panel). The horizontal axis is the
    separation from Uranus in planetary radii, and the vertical axis
    shows the percentage of the total dynamical simulation that
    particles survived at that location, represented by the green
    line. The vertical dotted lines represent the semi-major axes of
    the major moons, and the vertical dashed lines represent the
    locations of the rigid and fluid Roche limits. The bottom panel
    shows the change in eccentricity that occurs for each particle as
    a function of their initial semi-major axis during the course of
    the $10^7$ year simulations.}
  \label{fig:sim}
\end{figure*}

The results of the simulations are summarized in the plots shown in
Figure~\ref{fig:sim}. The top two panels show green lines that
represent the survival time of the injected particles (as a percentage
of the total simulation time) as a function of semi-major axis (in
units of planetary radii). The results after $10^6$ years and $10^7$
years and shown in the top and middle panels, respectively. The
vertical dotted lines indicate the locations of the major moons, and
the vertical dashed lines represent the locations of the rigid and
fluid body Roche limits for Uranus. A comparison of the top and middle
panels show that much of the particle loss has transpired by $10^6$
years. The peaks at each of the moon locations is evidence of their
ability to harbor trojan bodies within their orbit. The exception to
this are Titania and Oberon, the most massive of the moons, whose
orbits are largely cleared by the time $10^7$ years has elapsed,
though Trojan bodies may remain. Of particular note is the relatively
large gap in simulation survival for the injected particles within the
range 15--25 Uranus radii. It is also worth noting that the inner
boundary of dynamical stability is primarily located at 4.24 Uranus
radii for the $10^7$ year simulation, which is just slightly exterior
to the extent of the $\nu$ ring (see Section~\ref{arch}). Therefore,
the dynamical constraints imposed by the major moons allow for an age
of the Uranus rings that is larger than $10^7$~years, though there are
numerous non-gravitational forces that can limit the lifetime of such
structures \citep{burns1979c}.

\begin{deluxetable*}{lrrrrrrrrrrrrr}
  \tablecolumns{14}
  \tablewidth{0pc}
  \tablecaption{\label{tab:mmr} Mean motion resonance locations for
    Uranus moons.}
  \tablehead{
    \colhead{Moon} &
    \colhead{$a$} &
    \colhead{3:1} &
    \colhead{5:2} &
    \colhead{7:3} &
    \colhead{2:1} &
    \colhead{3:2} &
    \colhead{7:5} &
    \colhead{5:7} &
    \colhead{2:3} &
    \colhead{1:2} &
    \colhead{3:7} &
    \colhead{2:5} &
    \colhead{1:3}
  }
  \startdata
  Miranda &  5.062 &  2.43 &  2.75 &  2.88 &  3.19 &  3.86 &  4.05 &  6.34 &  6.63 &  8.04 &  8.91 &  9.33 & 10.53 \\
  Ariel   &  7.474 &  3.59 &  4.06 &  4.25 &  4.71 &  5.70 &  5.97 &  9.35 &  9.79 & 11.86 & 13.15 & 13.77 & 15.55 \\
  Umbriel & 10.419 &  5.01 &  5.66 &  5.92 &  6.56 &  7.95 &  8.33 & 13.04 & 13.65 & 16.54 & 18.33 & 19.19 & 21.67 \\
  Titania & 17.055 &  8.20 &  9.26 &  9.69 & 10.74 & 13.02 & 13.63 & 21.34 & 22.35 & 27.07 & 30.00 & 31.42 & 35.48 \\
  Oberon  & 22.830 & 10.98 & 12.39 & 12.98 & 14.38 & 17.42 & 18.24 & 28.57 & 29.92 & 36.24 & 40.16 & 42.05 & 47.49 \\
  \enddata
  \tablecomments{All distances are in units of Uranus radii.}
\end{deluxetable*}

As described in Section~\ref{moons}, a major contributor to the
dynamics of the system are the locations of MMR caused by the
gravitational influence of the major moons. Table~\ref{tab:mmr}
provided the semi-major axis, $a$, and the important MMR locations for
each of the major moons in units of Uranus radii. For example, the
$10^7$ year simulation shows a line of instability located 3.59 Uranus
radii, which corresponds to the location of the 3:1 MMR with Ariel,
and creates a dynamically unstable region within the ring
structure. In contrast, a spike of stability is present at $10^7$
years located at 20.1 Uranus radii. According to Table~\ref{tab:mmr},
this does not correspond to a location of MMR, but rather is
surrounded by other regions of MMR caused by the moons Umbriel,
Titania, and Oberon. As one final example, a location of instability
is apparent in both the $10^6$ year and $10^7$ year simulations
results at 13.02 Uranus radii, which corresponds to the 3:2 MMR with
Titania and closely aligns with the 5:7 MMR with Umbriel.

Even when injected particles are not lost from the system, the
gravitational influence of the moons may still manifest by altering
the orbits of the particles. The bottom panel of Figure~\ref{fig:sim}
shows the change in eccentricity, $\delta e$, that occurs for each
particle at the end of the $10^7$ year simulation. For example, the
top and middle panels of Figure~\ref{fig:sim} indicate a complicated
structure of Titania and Oberon MMR locations between 25--29 Uranus
radii, beyond which the gravitational perturbation effects cease to
have a significant impact on simulation survival rates. However, the
bottom panel of Figure~\ref{fig:sim} reveals that the extensive MMR
locations beyond 29 Uranus radii (see Table~\ref{tab:mmr}) result in
the excitation of particles into eccentric orbits. For example, the
location of the 3:7 MMR with Titania and 2:3 MMR with Oberon combine
to cause a significant region of higher eccentricity at $\sim$30
Uranus radii. Furthermore, the locations of 3:1 and 5:2 MMR with Ariel
excite particles between 3--4 Uranus radii. For such small separations
from the giant planet, tidal dissipation and circularization become
important long-term effects \citep{ogilvie2014a}, especially as
particles beyond the Roche limit accrete into moonlets (see
Section~\ref{discussion}). Since tidal dissipation is not included in
our simulations, the eccentricities shown in the bottom panel of
Figure~\ref{fig:sim} may be considered upper limits on the
perturbative effects of the moons on the injected particles. It should
further be noted that MMR locations in isolation do not intrinsically
result in instability, but rather the overlap of nonlinear secular
resonances can serve as a driveer for secular chaos
\citep{lithwick2011b}. Thus, it is the complex combination of the
Uranian moon MMR locations that produce much of the observed mass loss
within the system.

\begin{figure*}
  \begin{center}
    \includegraphics[angle=270,width=16.0cm]{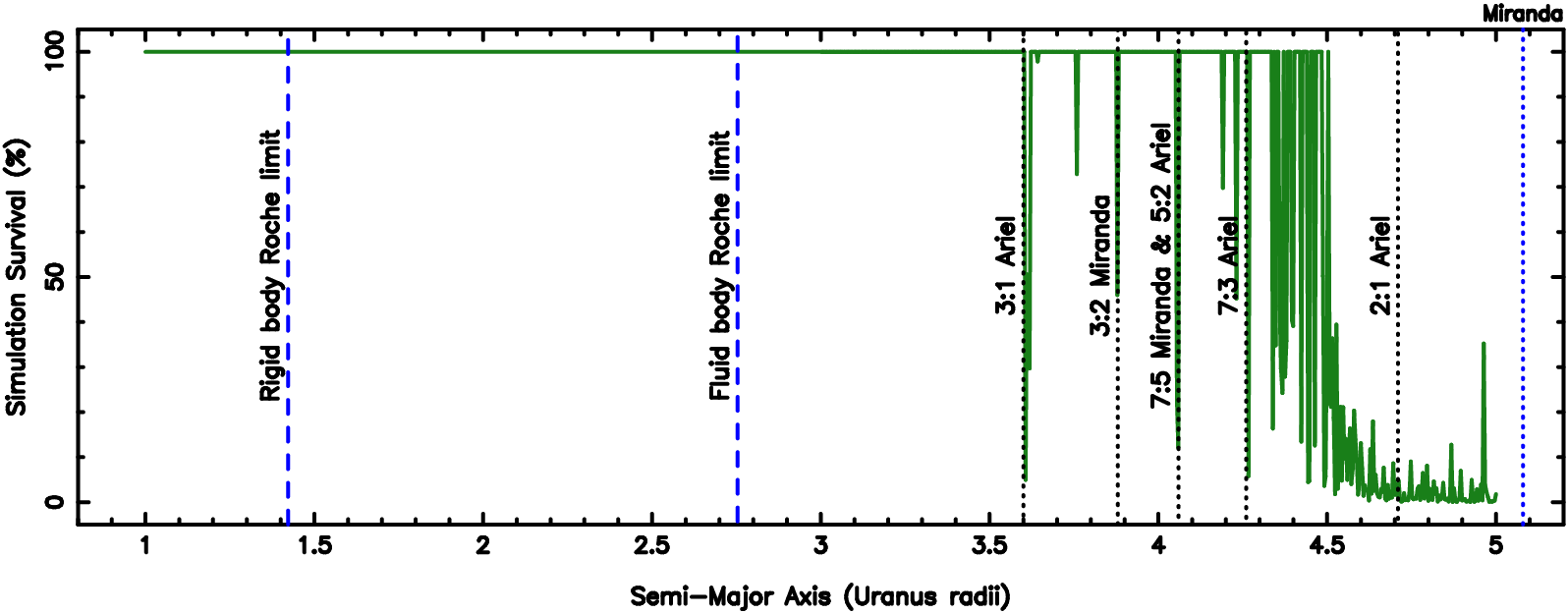}
  \end{center}
  \caption{Results of the $10^7$ year particle injection at higher
    resolution in the semi-major axis range of 1--5 Uranus radii. As
    for the top two panels of Figure~\ref{fig:sim}, the green line
    represents the percentage survival of the injected particle, the
    vertical dashed lines represent the Roche limit locations, and the
    blue vertical dotted line represents the semi-major axis of
    Miranda. The black vertical dotted lines indicate the locations of
    some MMR locations, corresponding to those values provided in
    Table~\ref{tab:mmr}.}
  \label{fig:finegrid}
\end{figure*}

To further explore the nature of the gravitational perturbations
taking place close to the planet, we perfomed an additional $10^7$
year simulation at higher resolution within the semi-major axis range
1--5 planetary radii. As for the previous simulations, we used 1000
evenly spaced semi-major axis locations which, in this case, produced
a distance resolution of $\sim$100~kms. The results of this simulation
are shown in Figure~\ref{fig:finegrid} where, in addition to the
features of Figure~\ref{fig:sim}, we include black vertical dotted
lines to indicate the MMR locations of greatest influence within the
1--5 planetary radii range. The increased resolution of this
simulation reveals important resonance features, including the
previously noted 3:1 MMR with Ariel. In particular, the effects of the
7:3 MMR with Ariel and the 3:2 MMR with Miranda are clearly visible,
as well as the coincident location of the 7:5 MMR with Miranda and 5:2
MMR with Ariel. The primary MMR within the region interior to Miranda
is the 2:1 MMR with Ariel, whose presence (combined with the 3:1 MMR
with Umbriel) plays a major role in clearing material between 4.5 and
5 planetary radii.

\begin{figure}
  \includegraphics[angle=270,width=8.5cm]{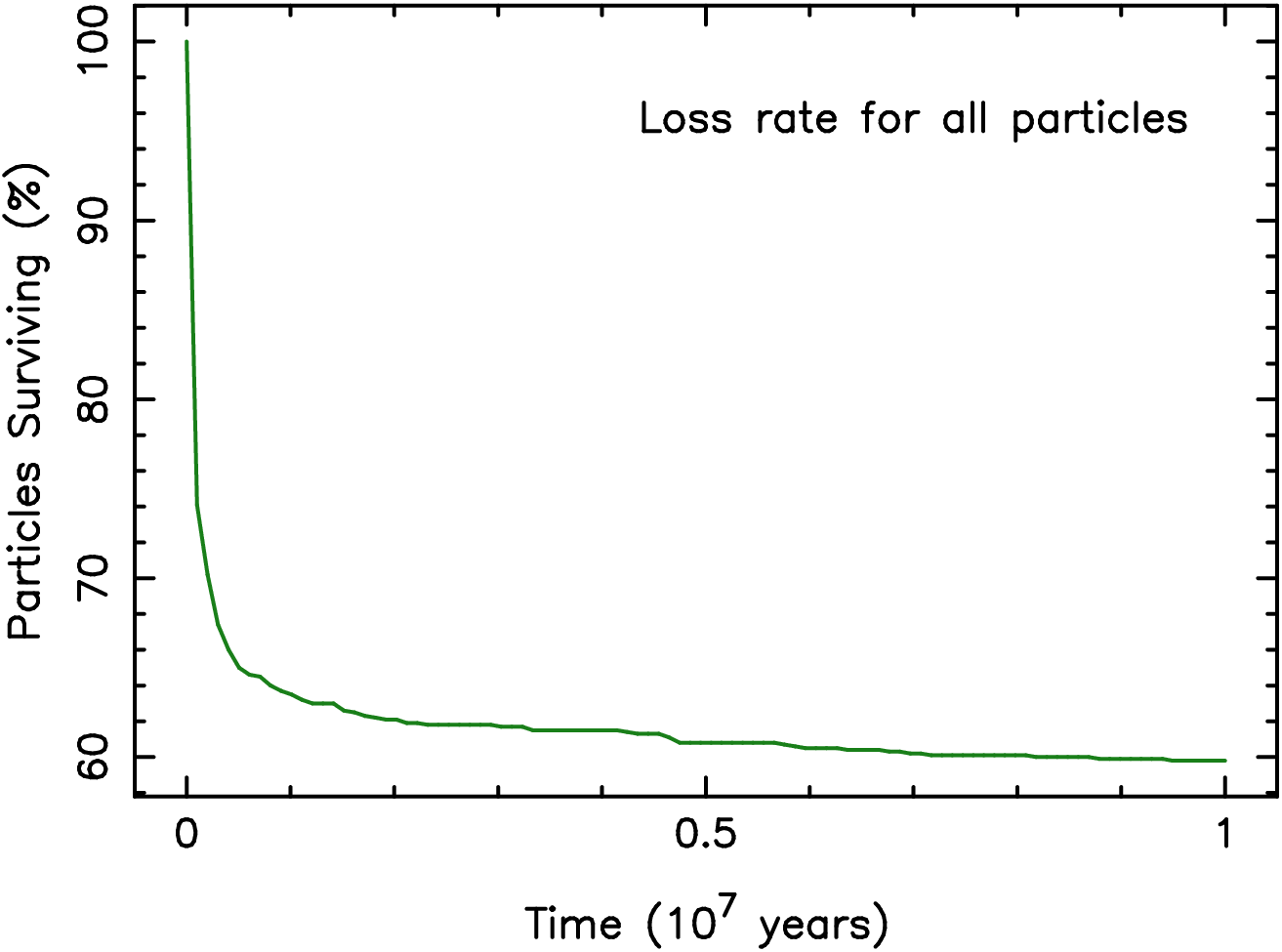}
  \caption{Loss rate for the injected particles into the Uranus
    system, shown as a function of time (fractions of $10^7$~years)
    and the percentage of particles that survive.}
  \label{fig:lossrate}
\end{figure}

As pointed out earlier, the comparison of the top and middle panels of
Figure~\ref{fig:sim} indicate that most of the mass loss occurred
within the initial $10^6$ year time frame. We investigated the mass
loss rate by calculating the percentage of particles surviving
throughout the full $10^7$ year simulation, the results of which are
shown in Figure~\ref{fig:lossrate}. These calculations reveal the
dramatic decline in the injected material into the system, showing
that 35\% of the total number of particles are lost within the first
$0.5 \times 10^6$~years of the simulation. Subsequent mass loss rates
are far more gradual, with an additional $\sim$5\% mass loss over the
following $\sim$$10^7$ years. Thus, once the full simulation time of
$10^7$ is reached, $\sim$40\% of the injected mass has been
lost. Therefore, there is a $\sim$13\% difference between the
remaining particle mass represented in the top two panels of
Figure~\ref{fig:sim}. The mass loss rates described here demonstrate
the rapid sculpting of material that occurs due to the gravitational
influence of the major moons orbiting Uranus, and that any rings that
form in the affected regions shown in Figure~\ref{fig:sim} are
relatively short-lived.


\section{Discussion}
\label{discussion}

The giant planets of the solar system have greatly influenced the
architecture and evolution of the inner solar system, as well as the
distribution of material and minor bodies out into the Kuiper belt
\citep{laskar1988b,deienno2011,dawson2012a,deienno2014,horner2020a,clement2021a,vervoort2022,kane2023a}.
A consequence of these interactions are the occurrence rate of impacts
that can be the source of material that remains distributed within the
Hill radius of the planet. Such material may be sourced from minor
bodies or moons that move within the Roche limit of the planet
\citep{canup1995,hyodo2017a,hyodo2017b}, from impactors on the moons
within the system
\citep{plescia1985,zahnle2003a,kirchoff2010,hueso2018b,ferguson2020},
or even moon collisions \citep{barbara2002,french2012c}. As described
in Section~\ref{intro}, impacts have played a profound role in the
evolution of Uranus, both intrinsically and to its dynamical
environment \citep{parisi2011b,reinhardt2020}, and may have been a
source for building the major moons of the Uranus system
\citep{salmon2022a}.

As described in Section~\ref{intro}, the ring system of Uranus is both
fascinating and complex
\citep{goldreich1979b,esposito1989,showalter2006a,ahearn2022}, and
whose evolution is likely connected to the simultaneous evolution of
the moons within the system
\citep{goldreich1987,porco1987a,canup1995}. Section~\ref{arch}
discusses the size of the Uranuian ring system, where tenuous rings
extend beyond the fluid Roche limit to $\sim$4 Uranus
radii. Contributions to the ring material may include collision
events, grinding down of small moons, and outgassing from the major
moons \citep{esposito2002}. Sustaining ring material within the
gravitational influence of the planet is subject to a variety of loss
processes, of which the dynamical effects caused by the major moons
are herein presented. The results of our simulations provided in
Section~\ref{influence} show the important consequence of these
dynamical effects on the sustainability of material within the system,
and the critical role of MMR locations in this context. Gravitational
dessication of ring material interior to the orbit of Miranda is
dominated by the Miranda and Ariel MMR locations, with a particularly
severe truncation of stable ring orbits occurring beyond $\sim$4.3
Uranus radii in the vicinity of Miranda and within the 2:1 MMR with
Ariel. Ringlets are sustainable in the gaps between moons out as far
as Titania, although such ring material would likely require being
sourced via a substantial impact or primordial formation material. Our
mass loss simulations show that the vast majority of coplanar material
is lost within a million years of orbital insertion, restricting and
scuplting the range of possible ring architectures for the system. The
mass loss caused by the major moons may have been further exacerbated
by previous orbital configurations if MMR crossing events between
moons did indeed occur \citep{tittemore1990a,cuk2020b}.

Beyond the dynamical loss processes caused by the major moons
described in this work, there are numerous other factors that can
desiccate material within the system, such as Poynting-Robertson drag
and the Yarkovsky effect \citep{rubincam2006,kobayashi2009a}.
However, many of these non-gravitational forces act upon material that
is significantly smaller than that considered in our simulations
\citep{burns1979c}. The bottom panel of Figure~\ref{fig:sim}
demonstrates the excitation of orbital eccentricities for particles
near locations of MMR. Such excitation beyond the Roche limit can
promote collisions and the accretion of material into
moonlets. Moonlet formation represents another means of mass loss from
the total distribution of orbiting particles, and can occur on
timescales that are comparable to the simulations described in this
work \citep{crida2012}. Another caveat to our methodology that is
worth noting is the ejection of particles into the equatorial plane of
the planet, particularly given the high axial tilt of the planet
relative to the orbital plane. Although the distribution of impactor
trajectories is unlikely to be isotropic, neither will they
necessarily be aligned with the present tilt of the Uranian rotational
axis. However, orbital precession of material that originates from
moon impacts or moon-moon collisions will result in an eventual
collapse of the material into the planet's equatorial plane, where
they will be subjected to the gravitational influences described in
this work. The high obliquity of Uranus does result in greater
exposure of the ring surface area to the solar wind
\citep{curtis1985e}. The effects of such exposure are largely
mitigated by the Uranus magnetic field
\citep{connerney1987b,stanley2006}, whose maximum strength occurs at
$\sim$4.2 Uranus radii, slightly beyond the extent of the rings, and
whose magnetopause boundary occurs at $\sim$18.0 Uranus radii,
slightly beyond the orbit of Titania \citep{ness1986}.


\section{Conclusions}
\label{conclusions}

The formation and evolution of the solar system giant planets and
their local environments is an extraordinarily complex research area,
with a vast number of associated interacting bodies. Each giant planet
system has a different story to tell, a story that is largely informed
by the properties of the planet (interior, atmosphere, orbit, spin
angular momentum), moons (composition, craters, geology, orbit), and
rings (composition, density, sustainability). The case of Uranus is of
particular interest since it has significant divergence in many of
these properties from its sibling giant planets, leading to
investigations on how aspects such as the high obliquity, ring
structure, and moon geology may be related to each other. A major step
in tracing such connections and their evolution is the concise
evaluation of the processes that are contributing and removing
material from the system that may yield insight into the plethora of
processes acting upon the material and their sources.

The simulation results presented here explore the specific component
of dynamical interactions within the Uranian system and their
contribution to mass loss processes. The five major moons of Uranus
are the major perturbative agents, and have proceeded through their
own complicated interactions including periods of MMR with each
other. Assuming dense material such as that described in
Section~\ref{influence}, our results predict a 35\% mass loss of
material, evenly distributed between 0--40 Uranus radii, within the
relatively small timeframe of $0.5 \times 10^6$ years. The vast
majority of the mass loss occurs within the semi-major axis range of
15--25 Uranus radii, where the perturbations are largely caused by the
major MMR locations with the two most massive moons of Titania and
Oberon. Their larger orbital distances from Uranus also enable them to
have larger Hill radii, further increasing their perturbative
influence over the injected particles. An additional region of mass
loss worth highlighting is that interior to the orbit of Miranda,
where the present ring system lies. The MMR locations resulting from
Miranda and Ariel dominate the mass loss within this region, and past
MMR crossings may have enhanced this mass loss effect. These regions
of significant particle loss can greatly truncate the sustainability
of rings or moonlet formation within the system.

The dynamical results presented here may be combined with moon
formation processes, the effects of orbital migration within an
accretion disk, and the non-gravitational forces acting upon the
injected material. However, a thorough analysis of the dynamics within
the Uranian system requires further data that is currently unavailable
for the planet and its moons. The interior structure of the planet,
its complex magnetic field, and the diverse geology of the moons, all
form the basis of key scientific questions that would be primarily
addressed via in-situ measurements from a Uranus orbitor mission
\citep{arridge2012,arridge2014,fletcher2020d,fletcher2020e,blanc2021,cartwright2021,fletcher2022}. Furthermore,
although their detection remains challenging with most current
detection techniques \citep{kane2011d}, ice giant analogs in
exoplanetary systems will aid in addressing these questions
\citep{wakeford2020b}. Overcoming the detection challenges will
provide a vast inventory of planets from which statistical studies of
occurrence rates, dynamical properties, and formation scenarios for
ice giant planets may be inferred. The increasing consolidation of
data from solar system and exoplanetary science will continue to
unveil the properties of ice giants \citep{horner2020b,kane2021d}, and
allow a deeper exploration of their dynamical histories.


\section*{Acknowledgements}

The authors would like to thank the anonymous referees for their
useful contributions and feebdack that improved the manuscript. The
results reported herein benefited from collaborations and/or
information exchange within NASA's Nexus for Exoplanet System Science
(NExSS) research coordination network sponsored by NASA's Science
Mission Directorate.


\software{REBOUND \citep{rein2012a}}




\end{document}